\documentclass[aps,prl,twocolumn,superscriptaddress]{revtex4}

\usepackage{graphicx}% Include figure files
\usepackage{graphics}% Include figure files
\usepackage{bm}
\begin{document}

\title{Identification of rolling resistance as a shape parameter in sheared granular media}

\author{Nicolas Estrada}\email[]{n.estrada22@uniandes.edu.co}
\affiliation{Departamento de Ingenier\'ia Civil y Ambiental - CeiBA Complex Systems Research Center, Universidad de Los Andes, Bogot\'a, Colombia}
\author{Emilien Az\'ema}\email[]{emilien.azema@univ-montp2.fr}
\author{Farhang Radjai}
\affiliation{LMGC,Universit\'e Montpellier 2 - CNRS, Place Eug\`ene Bataillon, 34095 Montpellier cedex 5, France}
\author{Alfredo Taboada}
\affiliation{Laboratoire G\'eosciences Montpellier, UMR 5243, Universit\'e Montpellier 2 - CNRS, Place Eug\`ene Bataillon, 34095 Montpellier cedex 5, France}

\date{\today}

\begin{abstract}
Using contact dynamics simulations, we compare the effect of rolling resistance at the contacts in granular systems composed of disks with the effect of angularity in granular systems composed of regular polygonal particles.
In simple shear conditions, we consider four aspects of the mechanical behavior of these systems in the steady state: shear strength, solid fraction, force and fabric anisotropies, and probability distribution of contact forces.
Our main finding is that, based on the energy dissipation associated with relative rotation between two particles in contact, the effect of rolling resistance can explicitly be identified with that of the number of sides in a regular polygonal particle.
This finding supports the use of rolling resistance as a shape parameter accounting for particle angularity and shows unambiguously that one of the
main influencing factors behind the mechanical behavior of granular systems composed of noncircular particles is the partial hindrance of rotations 
as a result of angular particle shape.
\end{abstract}

%\keywords{}

\maketitle

Most numerical studies on the mechanical behavior of granular materials deal with model systems composed of disks in 2D or spheres in 3D. 
This is usually due to the technical difficulties that arise when dealing with particles of complex shapes in experiments or discrete element methods. 
However, real granular materials are rarely composed of spherical particles, and it has been shown that the nonspherical shape of the grains strongly influences the mechanical behavior of granular systems.
This influence can be evidenced when characterizing the shear strength \cite{Ting1993,Ouadfel2001,NouguierLehon2003,Azema2007} and solid fraction \cite{Ting1993,Donev2004a,Man2005,Jiao2010}, as well as microstructural properties such as the distribution of contact forces \cite{CruzHidalgo2009,Azema2009}.
The effect of grain shape is thus a crucial aspect to be taken into account for a realistic description of granular systems.

One of the numerical ``tricks'' that can be used to obtain realistic values of strength and solid fraction while using only circular particles in simulations is to partially restrict the relative rotations between grains \cite{Ai2011}.
For example, several studies have shown that rolling resistance leads to shear strengths and solid fractions that are comparable to those observed in granular soils and rocks, e.g., \cite{Iwashita1998, Delenne2004, Calvetti2005a, Jiang2006}.
However, the extent to which rolling resistance can actually be compared to angular shape in more general terms, or whether rolling resistance and angular shape lead to similar structures at the mesoscopic scale, are interesting issues that remain poorly understood.

In this Letter, we compare, by means of discrete element simulations, the effects of rolling resistance and angularity.
We construct two sets of polydisperse 2D packings.
In the first set, the packings are composed of disks with an increasing magnitude of rolling resistance, whereas in the second set, the packings are composed of regular polygonal particles of increasing number of sides.
By comparing various properties extracted from the two sets, we find a remarkable matching of the data from the disk packings with those of the polygon packings for a rolling resistance expressed by a simple equation as a function of the number of sides.
This one-to-one mapping between the two sets is based on energy dissipation considerations and might be generalized to other particle shapes.

All packings are made up of $7500$ grains with diameters uniformly distributed by volume fractions between $0.6d$ and $2.4d$, where $d$ is the mean diameter. 
In all simulations, the coefficient of sliding friction $\mu_s$ between particles is $0.4$ and collisions are perfectly inelastic.
The particles are initially placed in a semiperiodic box $100d$ wide using a geometrical procedure \cite{Taboada2005}.
Next, the packing is sheared by imposing a constant shear velocity and a constant confining stress.
To avoid strain localization at the boundaries, sliding and rolling are inhibited for the particles in contact with the walls.
The samples are sheared up to a large cumulative shear strain $\gamma= \Delta x/ h = 5$, where $\Delta x$ is the horizontal displacement of the upper wall and $h$ is the thickness of the sample.
All measures are averaged over the last $50\%$ of cumulative shear strain in order to guarantee that they characterize the behavior of the system in the steady state, also known as the ``critical state'' in soil mechanics.
In all tests, the gravity is set to zero.

The simulations were carried out by means of the contact dynamics 
method \cite{Moreau1994,Jean1995,Jean1999,Radjai2009}, which assumes perfectly rigid particles interacting through mutual exclusion and Coulomb friction.
For specific implementation of the contact dynamics method see \cite{Taboada2005,Radjai2009}.

In the first set of samples, composed of disks, the rolling resistance is introduced through a {\em rolling friction law} \cite{Estrada2008}, analogous 
to the sliding friction law. 
Although rolling friction is introduced here as a numerical parameter, 
it may reflect various material parameters such as hysteresis, 
micro-sliding when the elastic moduli are different, inelasticity (in particular for polymers) and surface roughness \cite{Johnson1999}.
This law assumes that a contact can transmit a torque $M$ not exceeding 
a limit value $M_{max}=\mu_r \ell f_n$, where $\mu_r$ is the coefficient of rolling friction, $\ell$ is the magnitude of the branch vector joining the centers of the contacting particles, and $f_n$ is the normal force.
The scaling of $M_{max}$ with $\ell$ is meant to make $\mu_r$ dimensionless.
Relative rotation between two grains in contact is allowed only if $M=M_{max}$. 

In the second set of samples, composed of regular polygonal particles, two types 
of contact may occur: (1) between a corner and a side, and (2) between two sides.
Side/side interactions represent two constraints and are treated by associating two contact points along the common side and applying the volume exclusion and the sliding friction law to each of them.
Thus, in practice, two contact forces are calculated at each side/side contact.
However, only their resultant and application point are physically relevant, and the result is independent of the choice of the two contact points \cite{Saussine2006, Note_LMGC90}.

The stress components can be calculated from the simulation 
data by the relation $\sigma_{ij}=n_c\langle f_i^c \ell_j^c\rangle$, where $n_c$ is the number of contacts per unit volume and the average runs over the contacts $c$ with contact force $f^c$ and branch 
vector $\ell^c$ \cite{Radjai1998}.
The mean stress is $p=(\sigma_1 + \sigma_2)/2$, where $\sigma_1$ and $\sigma_2$ are the principal stress values, and the deviatoric stress is $q=(\sigma_1 - \sigma_2)/2$.
It is worth noting that in the presence of rolling resistance the stress tensor 
can be asymmetric and a couple-stress tensor may be 
added to the description. However, in all our tests the 
asymmetry is  negligibly small (i.e., $|(\sigma_{12}-\sigma_{21})/(\sigma_{12}+\sigma_{21})|\leq 0.0002$), suggesting that we do not need to consider the couple stress in the present study. 
Similar observations on the contribution of the couple stress tensor are reported in \cite{Oda2000}.

\begin{figure}
\includegraphics[width=7cm]{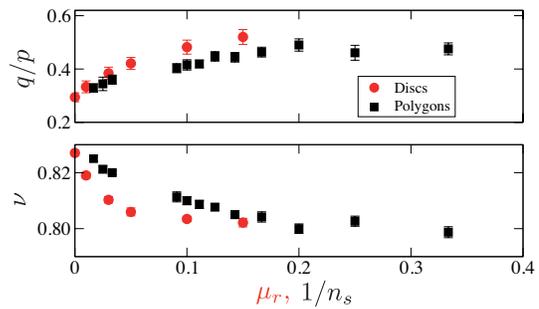}
\caption{\label{Fig:2a} (color online) Shear strength $q/p$ (Up) and 
solid fraction $\nu$ (Down) as functions of $\mu_r$ for the disks and of $1/n_s$ for the polygons. Error bars indicate the standard deviation.}
\end{figure}

Figure \ref{Fig:2a} shows the shear strength $q/p$ and solid fraction $\nu=V_p/V$, where $V_p$ is the volume occupied by the particles and $V$ is the total volume, 
as functions of $\mu_r$ for the disks and of $1/n_s$ for the polygons, where $n_s$ is the number of sides of the polygons.
It can be seen that both $q/p$ and $\nu$ follow similar trends in the two sets as $\mu_r$ and $1/n_s$ increase.
However, a direct comparison of the data between the two sets is not possible in this representation due to the different physical meanings of $\mu_r$ and $1/n_s$. 

\begin{figure}
\includegraphics[width=5.5cm]{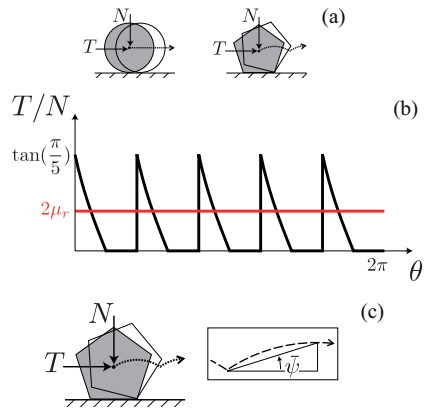}
\caption{\label{Fig:1} (a) Schema of rolling on a plane.
(b) Horizontal force $T$ required for rolling a distance equal to the perimeter.  
(c) Trajectory of the center of mass of the polygon (dashed line) and definition of the mean dilatancy angle $\bar{\psi}$.}
\end{figure}

The respective effects of rolling friction and angular shape can be compared by their roles in the hindering of relative rotation. 
Let us consider a particle (a disk with rolling friction and a regular pentagon) that rolls on a horizontal plane with a vertical force $N$ exerted at its center of mass and that is not allowed to slide; see Fig. \ref{Fig:1}(a).
Figure \ref{Fig:1}(b) shows the horizontal force $T$ that must be applied at the center of mass in order to make the particle roll, as a function of the rotation angle $\theta$.
The work needed to displace the particle a distance equal to its perimeter is
\begin{equation}
\label{Eq:1}
W_d=4\pi\mu_rR_dN
\end{equation}
for the disk with rolling friction, where $R_d$ is the radius of the disk and the magnitude of the branch vector $\ell$ (necessary to calculate $M_{max}$) has been replaced by the disk diameter, and
\begin{equation}
\label{Eq:2}
W_p=n_s(1-\cos(\pi/n_s))R_pN
\end{equation}
for the polygon, where $R_p$ is the radius of its circumcircle. 
Assuming equal work, i.e. $W_d = W_p$, we arrive at the following mapping between $\mu_r$ and $n_s$
\begin{equation}
\mu_r=(1/4)\tan \bar{\psi},
\label{eqn:mur}
\end{equation}
where it has been assumed that both particles have the same perimeter (i.e., $R_p=R_d(\pi/n_s)/\sin(\pi/n_s)$), and $\bar{\psi}=\pi/(2n_s)$ is the mean dilatancy angle of the trajectory of the center of mass of the polygon (see Fig.  \ref{Fig:1}(c)).
For a similar attempt to quantify the role of grain shapes in hindering relative rotation, see \cite{Matsushima2005}.

\begin{figure}
\includegraphics[width=7cm]{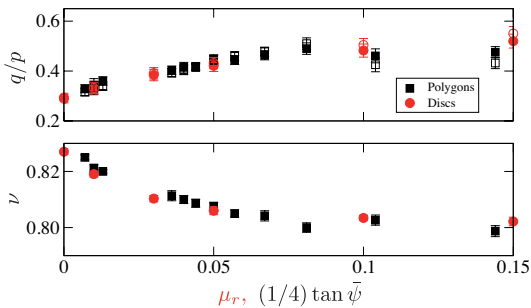}
\caption{\label{Fig:2b} (color online) (Up) Shear strength $q/p$ as a function of $\mu_r$ for the disks and of $(1/4)\tan \bar{\psi}$ for the polygons, both from raw simulation data (full symbols) and as predicted by Eq. \ref{Eq:HarmonicQP} (empty symbols). (Down) Solid fraction $\nu$ as a function of $\mu_r$ for the disks and of $(1/4)\tan \bar{\psi}$ for the polygons.
Error bars indicate the standard deviation.}
\end{figure}

Figure \ref{Fig:2b} shows the shear strength $q/p$ and solid fraction $\nu$ as functions of $\mu_r$ for the disks and of $(1/4)\tan \bar{\psi}$ for the polygons.
Remarkably, the shear strengths and solid fractions of the two sets of packings collapse, both increasing and decreasing, respectively, with $\mu_r$ and $(1/4)\tan \bar{\psi}$ and tending to a constant value at $\mu_r = (1/4)\tan \bar{\psi} \simeq 0.1$.
In other words, from a macro-scale viewpoint, a packing of regular polygons of $n_s$ sides is equivalent to a packing of disks with a coefficient of rolling friction $\mu_r$ given by Eq. \ref{eqn:mur}.
This result supports also the choice of the required energy for rolling as a relevant physical quantity for the rheology of granular materials.

\begin{figure}
\includegraphics[width=6.5cm]{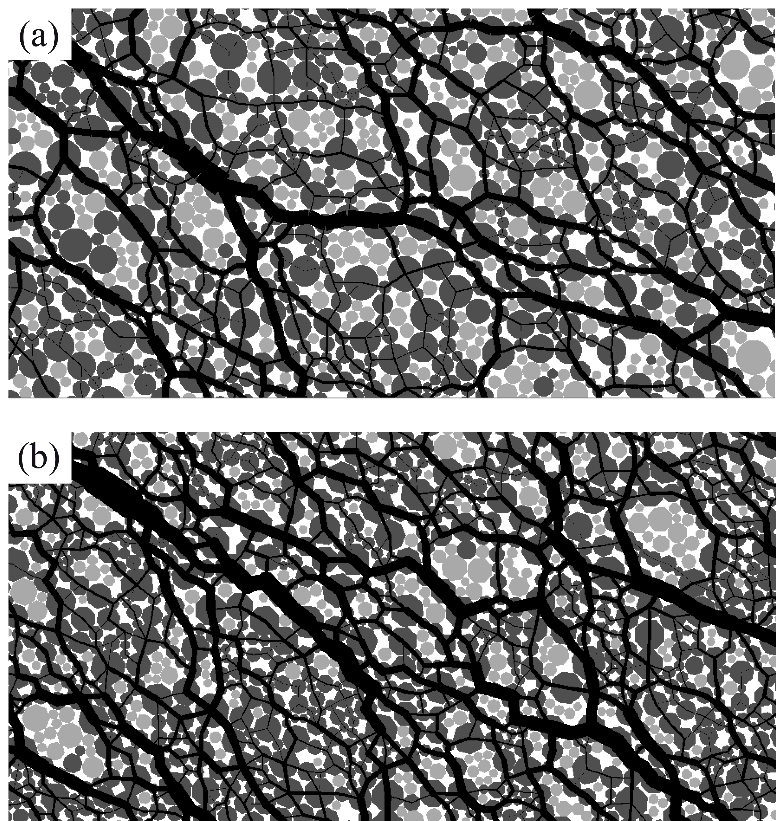}
\caption{\label{Fig:3} Snapshots of the force network in (a) a system composed of disks with rolling friction ($\mu_r=0.05$) and (b) a system composed of octagonal particles.
The line thickness is proportional to the normal force.
The floating particles are represented in light grey.}
\end{figure}

The mapping evidenced in Fig. \ref{Fig:2b} hints at similar packing structures in the two sets.
Figure \ref{Fig:3} shows two snapshots: one representing a disk packing with $\mu_r=0.05$ and the other representing a polygon packing with $n_s=8$ (note that $0.05 \simeq (1/4) \tan(\pi/(2*8))$ according to Eq. \ref{eqn:mur}).
The contact forces are represented by segments joining the particle centers, with a thickness proportional to the force magnitude. 
We observe that the force-carrying backbone is astonishingly similar in the two systems.

From the expression of the stress tensor, it can be shown that the shear stress $q/p$ reflects the packing structure and force transmission via a simple relation \cite{Rothenburg1989}:
\begin{equation}
q/p \simeq (1/2)(a_c + a_n + a_t),
\label{Eq:HarmonicQP}
\end{equation}
where $a_c$, $a_n$, and $a_t$,  are the anisotropies of the angular distributions of contact orientations $P_{n}(\theta)$, normal forces $\langle f_n \rangle(\theta)$, and tangential forces $\langle f_t \rangle(\theta)$, respectively, as a function of contact orientation $\theta$, which are approximated by their lowest order Fourier expansions:
\begin{eqnarray}
P_{ n}(\theta) &\simeq &1/\pi\{1 +a_c \cos2(\theta-\theta_c)\}, \nonumber \\
\langle f_n \rangle (\theta) & \simeq& \langle f_n \rangle \{1 +a_n \cos2(\theta-\theta_n)\},  \nonumber \\
\langle f_t \rangle (\theta) & \simeq& -\langle f_n \rangle a_t \sin(\theta-\theta_t),
\label{Eq:Aniso}
\end{eqnarray}
where  $\langle f_n \rangle$ is the mean normal fore, and $\theta_c = \theta_n = \theta_t$ are the corresponding privileged directions, which, in the steady state, coincide with the principal stress direction.
Equation \ref{Eq:HarmonicQP} reveals distinct origins of the shear strength in terms of force and texture anisotropy.
The empty symbols in Fig. \ref{Fig:2b} show $q/p$ as predicted by Eq. \ref{Eq:HarmonicQP}.
We see that this equation approximates well the shear strength for all raw data.

\begin{figure}
\includegraphics[width=7cm]{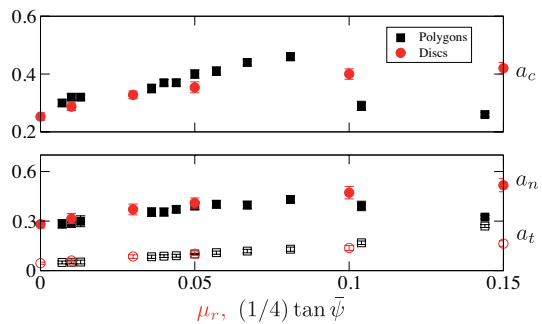}
\caption{\label{Fig:5} (color online) Contact anisotropy $a_c$ (Up) and force anisotropies, $a_n$ (full symbols) and $a_t$ (empty symbols) (Down), as functions of $\mu_r$ for the disks and of $(1/4)\tan \bar{\psi}$ for the polygons.
Error bars indicate the standard deviation.}
\end{figure}

The anisotropies $a_c$, $a_n$, and $a_t$ are shown in Fig. \ref{Fig:5} as functions of $\mu_r$ for the disks and of $(1/4)\tan \bar{\psi}$ for the polygons.
It is remarkable that all anisotropies are almost identical between the two sets.
This correspondence is only broken for polygons with small numbers of sides, i.e., for $n_s=3$ and $4$.
This happens because for these polygons the contact orientation is strongly influenced by 
the low rotational symmetry of the particles and the orientations of the sides rather than the relative positions of the particles.

\begin{figure}
\includegraphics[width=7cm]{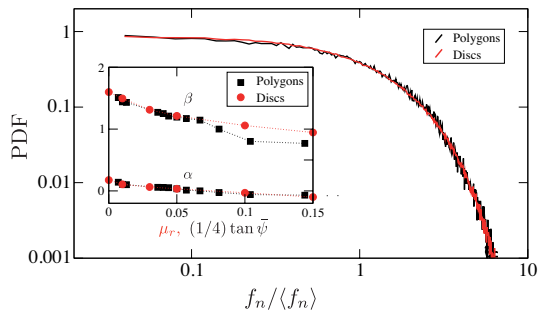}
\caption{\label{Fig:6} (color online) Probability distribution function of normalized normal forces $f_n/\langle f_n \rangle$ for the two systems of Fig. \ref{Fig:3}.
The inset shows the evolution of the exponents $\alpha$ and $\beta$ (see text for definitions) with $\mu_r$ for the disks and with $(1/4)\tan \bar{\psi}$ for the polygons.}
\end{figure}

The mapping between rolling friction and angular shape of particles is also reflected by the probability density function (PDF) of normal forces displayed in Fig. \ref{Fig:6}.
In this figure, we compare the PDFs of the two samples shown in Fig. \ref{Fig:3}.
The two PDFs are almost identical.
The PDF can be approximated by a power law (i.e., $PDF \propto \left({f_n}/{\langle f_n \rangle}\right)^{-\alpha}$) in the range of small forces and by an exponential function (i.e., $PDF \propto \ e^{\beta \left(1-{f_n}/{\langle f_n \rangle}\right)}$) in the range of strong forces \cite{Radjai1996}.
The inset in Fig. \ref{Fig:6} shows the coefficients $\alpha$ and $\beta$ as functions of $\mu_r$ for the disks and of $(1/4)\tan \bar{\psi}$ for the polygons, confirming that the similarity of the PDFs is maintained for the whole range of rolling frictions and numbers of sides studied in this work.

To sum up, the simulations presented in this Letter provide strong evidence for the mapping between the two studied parameters, i.e., rolling resistance and shape angularity.
This correspondence was established by considering shear strength, solid fraction, force and fabric anisotropies, and the PDFs of normal forces in the steady state. 
A practical consequence of this finding is that rolling resistance may be employed to imitate the effect of angular shape in discrete-element simulations of granular materials. 
More importantly, it suggests that the hindrance of particle rotations is a major effect of angular particle shape.
In this picture, the effect of rolling friction is to force the particles to rearrange as if they were glued to each other. In this way, the clusters of two or more particles behave as non-circular particles. 
This result may be tested in other loading conditions and it is potentially extensible to other particle shapes (in 2D and 3D), opening new scopes in modeling complex granular systems.

This work was financially supported by the France-Colombia Ecos-Nord project (Grant No. C08U01).

%\bibliography{references.bib}

\begin{thebibliography}{28}
\expandafter\ifx\csname natexlab\endcsname\relax\def\natexlab#1{#1}\fi
\expandafter\ifx\csname bibnamefont\endcsname\relax
  \def\bibnamefont#1{#1}\fi
\expandafter\ifx\csname bibfnamefont\endcsname\relax
  \def\bibfnamefont#1{#1}\fi
\expandafter\ifx\csname citenamefont\endcsname\relax
  \def\citenamefont#1{#1}\fi
\expandafter\ifx\csname url\endcsname\relax
  \def\url#1{\texttt{#1}}\fi
\expandafter\ifx\csname urlprefix\endcsname\relax\def\urlprefix{URL }\fi
\providecommand{\bibinfo}[2]{#2}
\providecommand{\eprint}[2][]{\url{#2}}

\bibitem[{\citenamefont{Ouadefl and Rothenburg}(2001)}]{Ouadfel2001}
\bibinfo{author}{\bibfnamefont{H.}~\bibnamefont{Ouadefl}} \bibnamefont{and}
  \bibinfo{author}{\bibfnamefont{L.}~\bibnamefont{Rothenburg}},
  \bibinfo{journal}{Mechanics of Materials} \textbf{\bibinfo{volume}{33}},
  \bibinfo{pages}{201} (\bibinfo{year}{2001}).

\bibitem[{\citenamefont{Nouguier-Lehon
  et~al.}(2003)\citenamefont{Nouguier-Lehon, Cambou, and
  Vincens}}]{NouguierLehon2003}
\bibinfo{author}{\bibfnamefont{C.}~\bibnamefont{Nouguier-Lehon}},
  \bibinfo{author}{\bibfnamefont{B.}~\bibnamefont{Cambou}}, \bibnamefont{and}
  \bibinfo{author}{\bibfnamefont{E.}~\bibnamefont{Vincens}},
  \bibinfo{journal}{Int. J. Numer. Anal. Meth. Geomech.}
  \textbf{\bibinfo{volume}{27}}, \bibinfo{pages}{1207} (\bibinfo{year}{2003}).

\bibitem[{\citenamefont{Az\'ema et~al.}(2007)\citenamefont{Az\'ema, Radja\"i,
  Peyroux, and Saussine}}]{Azema2007}
\bibinfo{author}{\bibfnamefont{E.}~\bibnamefont{Az\'ema}},
  \bibinfo{author}{\bibfnamefont{F.}~\bibnamefont{Radja\"i}},
  \bibinfo{author}{\bibfnamefont{R.}~\bibnamefont{Peyroux}}, \bibnamefont{and}
  \bibinfo{author}{\bibfnamefont{G.}~\bibnamefont{Saussine}},
  \bibinfo{journal}{Phys. Rev. E} \textbf{\bibinfo{volume}{76}},
  \bibinfo{pages}{011301} (\bibinfo{year}{2007}).

\bibitem[{\citenamefont{Ting et~al.}(1993)\citenamefont{Ting, Khwaja, Meachum,
  and Rowell}}]{Ting1993}
\bibinfo{author}{\bibfnamefont{J.}~\bibnamefont{Ting}},
  \bibinfo{author}{\bibfnamefont{M.}~\bibnamefont{Khwaja}},
  \bibinfo{author}{\bibfnamefont{L.}~\bibnamefont{Meachum}}, \bibnamefont{and}
  \bibinfo{author}{\bibfnamefont{J.}~\bibnamefont{Rowell}},
  \bibinfo{journal}{Int. J. Numer. Anal. Meth. Geomech.}
  \textbf{\bibinfo{volume}{17}}, \bibinfo{pages}{603 } (\bibinfo{year}{1993}).

\bibitem[{\citenamefont{Donev et~al.}(2004)\citenamefont{Donev, Stillinger,
  Chaikin, and Torquato}}]{Donev2004a}
\bibinfo{author}{\bibfnamefont{A.}~\bibnamefont{Donev}},
  \bibinfo{author}{\bibfnamefont{F.~H.} \bibnamefont{Stillinger}},
  \bibinfo{author}{\bibfnamefont{P.~M.} \bibnamefont{Chaikin}},
  \bibnamefont{and} \bibinfo{author}{\bibfnamefont{S.}~\bibnamefont{Torquato}},
  \bibinfo{journal}{Phys. Rev. Lett.} \textbf{\bibinfo{volume}{92}},
  \bibinfo{pages}{255506} (\bibinfo{year}{2004}).

\bibitem[{\citenamefont{Man et~al.}(2005)\citenamefont{Man, Donev, Stillinger,
  Sullivan, Russel, Heeger, Inati, Torquato, and Chaikin}}]{Man2005}
\bibinfo{author}{\bibfnamefont{W.}~\bibnamefont{Man}},
  \bibinfo{author}{\bibfnamefont{A.}~\bibnamefont{Donev}},
  \bibinfo{author}{\bibfnamefont{F.~H.} \bibnamefont{Stillinger}},
  \bibinfo{author}{\bibfnamefont{M.~T.} \bibnamefont{Sullivan}},
  \bibinfo{author}{\bibfnamefont{W.~B.} \bibnamefont{Russel}},
  \bibinfo{author}{\bibfnamefont{D.}~\bibnamefont{Heeger}},
  \bibinfo{author}{\bibfnamefont{S.}~\bibnamefont{Inati}},
  \bibinfo{author}{\bibfnamefont{S.}~\bibnamefont{Torquato}}, \bibnamefont{and}
  \bibinfo{author}{\bibfnamefont{P.~M.} \bibnamefont{Chaikin}},
  \bibinfo{journal}{Phys. Rev. Lett.} \textbf{\bibinfo{volume}{94}},
  \bibinfo{pages}{198001} (\bibinfo{year}{2005}).

\bibitem[{\citenamefont{Jiao et~al.}(2010)\citenamefont{Jiao, Stillinger, and
  Torquato}}]{Jiao2010}
\bibinfo{author}{\bibfnamefont{Y.}~\bibnamefont{Jiao}},
  \bibinfo{author}{\bibfnamefont{F.~H.} \bibnamefont{Stillinger}},
  \bibnamefont{and} \bibinfo{author}{\bibfnamefont{S.}~\bibnamefont{Torquato}},
  \bibinfo{journal}{Phys. Rev. E} \textbf{\bibinfo{volume}{81}},
  \bibinfo{pages}{041304} (\bibinfo{year}{2010}).

\bibitem[{\citenamefont{Cruz-Hidalgo et~al.}(2009)\citenamefont{Cruz-Hidalgo,
  Zuriguel, Maza, and Pagonabarraga}}]{CruzHidalgo2009}
\bibinfo{author}{\bibfnamefont{R.}~\bibnamefont{Cruz-Hidalgo}},
  \bibinfo{author}{\bibfnamefont{I.}~\bibnamefont{Zuriguel}},
  \bibinfo{author}{\bibfnamefont{D.}~\bibnamefont{Maza}}, \bibnamefont{and}
  \bibinfo{author}{\bibfnamefont{I.}~\bibnamefont{Pagonabarraga}},
  \bibinfo{journal}{Phys. Rev. Lett.} \textbf{\bibinfo{volume}{103}},
  \bibinfo{pages}{118001} (\bibinfo{year}{2009}).

\bibitem[{\citenamefont{Az\'ema et~al.}(2009)\citenamefont{Az\'ema, Radjai, and
  Saussine}}]{Azema2009}
\bibinfo{author}{\bibfnamefont{E.}~\bibnamefont{Az\'ema}},
  \bibinfo{author}{\bibfnamefont{F.}~\bibnamefont{Radjai}}, \bibnamefont{and}
  \bibinfo{author}{\bibfnamefont{G.}~\bibnamefont{Saussine}},
  \bibinfo{journal}{Mechanics of Materials} \textbf{\bibinfo{volume}{41}},
  \bibinfo{pages}{729} (\bibinfo{year}{2009}).

\bibitem[{\citenamefont{Ai et~al.}(2011)\citenamefont{Ai, Chen, Rotter, and
  Ooi}}]{Ai2011}
\bibinfo{author}{\bibfnamefont{J.}~\bibnamefont{Ai}},
  \bibinfo{author}{\bibfnamefont{J.-F.} \bibnamefont{Chen}},
  \bibinfo{author}{\bibfnamefont{J.~M.} \bibnamefont{Rotter}},
  \bibnamefont{and} \bibinfo{author}{\bibfnamefont{J.~Y.} \bibnamefont{Ooi}},
  \bibinfo{journal}{Powder Technology} \textbf{\bibinfo{volume}{206}},
  \bibinfo{pages}{269} (\bibinfo{year}{2011}).

\bibitem[{\citenamefont{Iwashita and Oda}(1998)}]{Iwashita1998}
\bibinfo{author}{\bibfnamefont{K.}~\bibnamefont{Iwashita}} \bibnamefont{and}
  \bibinfo{author}{\bibfnamefont{M.}~\bibnamefont{Oda}},
  \bibinfo{journal}{Journal of Engineering Mechanics}
  \textbf{\bibinfo{volume}{124}}, \bibinfo{pages}{285} (\bibinfo{year}{1998}).

\bibitem[{\citenamefont{Delenne et~al.}(2004)\citenamefont{Delenne,
  {El~Youssoufi}, Cherblanc, and B\'enet}}]{Delenne2004}
\bibinfo{author}{\bibfnamefont{J.-Y.} \bibnamefont{Delenne}},
  \bibinfo{author}{\bibfnamefont{M.~S.} \bibnamefont{{El~Youssoufi}}},
  \bibinfo{author}{\bibfnamefont{F.}~\bibnamefont{Cherblanc}},
  \bibnamefont{and} \bibinfo{author}{\bibfnamefont{J.-C.}
  \bibnamefont{B\'enet}}, \bibinfo{journal}{Int. J. Numer. Anal. Meth.
  Geomech.} \textbf{\bibinfo{volume}{28}}, \bibinfo{pages}{1577}
  (\bibinfo{year}{2004}).

\bibitem[{\citenamefont{Calvetti and Nova}(2005)}]{Calvetti2005a}
\bibinfo{author}{\bibfnamefont{F.}~\bibnamefont{Calvetti}} \bibnamefont{and}
  \bibinfo{author}{\bibfnamefont{R.}~\bibnamefont{Nova}}, in
  \emph{\bibinfo{booktitle}{Powders and Grains, 5th. International Conference
  on Micromechanics of Granular Media, Stuttgart}}, edited by
  \bibinfo{editor}{\bibfnamefont{H.~H.} \bibnamefont{R.~Garc\'ia~Rojo}}
  \bibnamefont{and} \bibinfo{editor}{\bibfnamefont{S.}~\bibnamefont{McNamara}}
  (\bibinfo{publisher}{A.A. Balkema}, \bibinfo{year}{2005}),
  vol.~\bibinfo{volume}{1}, pp. \bibinfo{pages}{245--249}.

\bibitem[{\citenamefont{Jiang et~al.}(2006)\citenamefont{Jiang, Yu, and
  Harris}}]{Jiang2006}
\bibinfo{author}{\bibfnamefont{M.~J.} \bibnamefont{Jiang}},
  \bibinfo{author}{\bibfnamefont{H.~S.} \bibnamefont{Yu}}, \bibnamefont{and}
  \bibinfo{author}{\bibfnamefont{D.}~\bibnamefont{Harris}},
  \bibinfo{journal}{Int. J. Numer. Anal. Meth. Geomech.}
  \textbf{\bibinfo{volume}{30}}, \bibinfo{pages}{723} (\bibinfo{year}{2006}).

\bibitem[{\citenamefont{Taboada et~al.}(2005)\citenamefont{Taboada, Chang,
  Radjai, and Bouchette}}]{Taboada2005}
\bibinfo{author}{\bibfnamefont{A.}~\bibnamefont{Taboada}},
  \bibinfo{author}{\bibfnamefont{K.-J.} \bibnamefont{Chang}},
  \bibinfo{author}{\bibfnamefont{F.}~\bibnamefont{Radjai}}, \bibnamefont{and}
  \bibinfo{author}{\bibfnamefont{F.}~\bibnamefont{Bouchette}},
  \bibinfo{journal}{J. Geophys. Res.} \textbf{\bibinfo{volume}{110}},
  \bibinfo{pages}{B09202} (\bibinfo{year}{2005}).

\bibitem[{\citenamefont{Moreau}(1994)}]{Moreau1994}
\bibinfo{author}{\bibfnamefont{J.~J.} \bibnamefont{Moreau}},
  \bibinfo{journal}{European Journal of Mechanics, A/Solids}
  \textbf{\bibinfo{volume}{13 (Suppl.)}}, \bibinfo{pages}{93}
  (\bibinfo{year}{1994}).

\bibitem[{\citenamefont{Jean}(1995)}]{Jean1995}
\bibinfo{author}{\bibfnamefont{M.}~\bibnamefont{Jean}},
  \emph{\bibinfo{title}{Mechanics of Geometrical Interfaces}}
  (\bibinfo{publisher}{Elsevier, New York}, \bibinfo{year}{1995}), pp.
  \bibinfo{pages}{463--486}.

\bibitem[{\citenamefont{Jean}(1999)}]{Jean1999}
\bibinfo{author}{\bibfnamefont{M.}~\bibnamefont{Jean}},
  \bibinfo{journal}{Comput. Methods Appl. Mech. Engrg}
  \textbf{\bibinfo{volume}{117}}, \bibinfo{pages}{235 } (\bibinfo{year}{1999}).

\bibitem[{\citenamefont{Radjai and Richefeu}(2009)}]{Radjai2009}
\bibinfo{author}{\bibfnamefont{F.}~\bibnamefont{Radjai}} \bibnamefont{and}
  \bibinfo{author}{\bibfnamefont{V.}~\bibnamefont{Richefeu}},
  \bibinfo{journal}{Mechanics of Materials} \textbf{\bibinfo{volume}{41}},
  \bibinfo{pages}{715 } (\bibinfo{year}{2009}).

\bibitem[{\citenamefont{Estrada et~al.}(2008)\citenamefont{Estrada, Taboada,
  and Radja\"i}}]{Estrada2008}
\bibinfo{author}{\bibfnamefont{N.}~\bibnamefont{Estrada}},
  \bibinfo{author}{\bibfnamefont{A.}~\bibnamefont{Taboada}}, \bibnamefont{and}
  \bibinfo{author}{\bibfnamefont{F.}~\bibnamefont{Radja\"i}},
  \bibinfo{journal}{Phys. Rev. E} \textbf{\bibinfo{volume}{78}},
  \bibinfo{pages}{021301} (\bibinfo{year}{2008}).

\bibitem[{\citenamefont{Johnson}(1999)}]{Johnson1999}
\bibinfo{author}{\bibfnamefont{K.}~\bibnamefont{Johnson}},
  \emph{\bibinfo{title}{Contact Mechanics}} (\bibinfo{publisher}{University
  Press, Cambridge}, \bibinfo{year}{1999}).

\bibitem[{\citenamefont{Saussine et~al.}(2006)\citenamefont{Saussine, Cholet,
  Gautier, Dubois, Bohatier, and Moreau}}]{Saussine2006}
\bibinfo{author}{\bibfnamefont{G.}~\bibnamefont{Saussine}},
  \bibinfo{author}{\bibfnamefont{C.}~\bibnamefont{Cholet}},
  \bibinfo{author}{\bibfnamefont{P.}~\bibnamefont{Gautier}},
  \bibinfo{author}{\bibfnamefont{F.}~\bibnamefont{Dubois}},
  \bibinfo{author}{\bibfnamefont{C.}~\bibnamefont{Bohatier}}, \bibnamefont{and}
  \bibinfo{author}{\bibfnamefont{J.}~\bibnamefont{Moreau}},
  \bibinfo{journal}{Comput. Methods Appl. Mech. Eng.}
  \textbf{\bibinfo{volume}{195}}, \bibinfo{pages}{2841 }
  (\bibinfo{year}{2006}).

\bibitem[{Not()}]{Note_LMGC90}
\bibinfo{note}{Polygonal particles are simulated with the LMGC90 platform
  developed in Montpellier by F. Dubois and M. Jean.}

\bibitem[{\citenamefont{Radjai et~al.}(1998)\citenamefont{Radjai, Wolf, Jean,
  and Moreau}}]{Radjai1998}
\bibinfo{author}{\bibfnamefont{F.}~\bibnamefont{Radjai}},
  \bibinfo{author}{\bibfnamefont{D.~E.} \bibnamefont{Wolf}},
  \bibinfo{author}{\bibfnamefont{M.}~\bibnamefont{Jean}}, \bibnamefont{and}
  \bibinfo{author}{\bibfnamefont{J.-J.} \bibnamefont{Moreau}},
  \bibinfo{journal}{Phys. Rev. Lett.} \textbf{\bibinfo{volume}{80}},
  \bibinfo{pages}{61} (\bibinfo{year}{1998}).

\bibitem[{\citenamefont{Oda and Iwashita}(2000)}]{Oda2000}
\bibinfo{author}{\bibfnamefont{M.}~\bibnamefont{Oda}} \bibnamefont{and}
  \bibinfo{author}{\bibfnamefont{K.}~\bibnamefont{Iwashita}},
  \bibinfo{journal}{International Journal of Engineering Science}
  \textbf{\bibinfo{volume}{38}}, \bibinfo{pages}{1713} (\bibinfo{year}{2000}).

\bibitem[{\citenamefont{Matsushima and Nova}(2005)}]{Matsushima2005}
\bibinfo{author}{\bibfnamefont{T.}~\bibnamefont{Matsushima}} \bibnamefont{and}
  \bibinfo{author}{\bibfnamefont{R.}~\bibnamefont{Nova}}, in
  \emph{\bibinfo{booktitle}{Powders and Grains, 5th. International Conference
  on Micromechanics of Granular Media, Stuttgart}}, edited by
  \bibinfo{editor}{\bibfnamefont{H.~H.} \bibnamefont{R.~Garc\'ia~Rojo}}
  \bibnamefont{and} \bibinfo{editor}{\bibfnamefont{S.}~\bibnamefont{McNamara}}
  (\bibinfo{publisher}{A.A. Balkema}, \bibinfo{year}{2005}),
  vol.~\bibinfo{volume}{1}, pp. \bibinfo{pages}{1319--1323}.

\bibitem[{\citenamefont{Rothenburg and Bathurst}(1989)}]{Rothenburg1989}
\bibinfo{author}{\bibfnamefont{L.}~\bibnamefont{Rothenburg}} \bibnamefont{and}
  \bibinfo{author}{\bibfnamefont{R.~J.} \bibnamefont{Bathurst}},
  \bibinfo{journal}{Geotechnique} \textbf{\bibinfo{volume}{39}},
  \bibinfo{pages}{601} (\bibinfo{year}{1989}).

\bibitem[{\citenamefont{Radjai et~al.}(1996)\citenamefont{Radjai, Jean, Moreau,
  and Roux}}]{Radjai1996}
\bibinfo{author}{\bibfnamefont{F.}~\bibnamefont{Radjai}},
  \bibinfo{author}{\bibfnamefont{M.}~\bibnamefont{Jean}},
  \bibinfo{author}{\bibfnamefont{J.-J.} \bibnamefont{Moreau}},
  \bibnamefont{and} \bibinfo{author}{\bibfnamefont{S.}~\bibnamefont{Roux}},
  \bibinfo{journal}{Phys. Rev. Lett.} \textbf{\bibinfo{volume}{77}},
  \bibinfo{pages}{274} (\bibinfo{year}{1996}).

\end{thebibliography}

\end{document}